\documentclass[sigconf]{acmart}
\usepackage{balance}

\usepackage{breakurl}

\renewcommand\footnotetextcopyrightpermission[1]{} 
\usepackage{cleveref}
\usepackage[]{ragged2e}

\usepackage{tcolorbox}
\usepackage[utf8]{inputenc} 
\usepackage{geometry} 
\usepackage{tikz} 
\usepackage{lipsum}
\usepackage{cleveref}
\crefname{section}{§}{§§}

\newcommand{\mybox}[4]{
    \begin{figure}[h]
        \centering
    \begin{tikzpicture}
        \node[anchor=text,text width=\columnwidth-1.2cm, draw, rounded corners, line width=1pt, fill=#3, inner sep=5mm] (big) {#4};
        \node[draw, rounded corners, line width=.5pt, fill=#2, anchor=west, xshift=5mm] (small) at (big.north west) {#1};
    \end{tikzpicture}
    \end{figure}
}
\AtBeginDocument{%
  \providecommand\BibTeX{{%
    \normalfont B\kern-0.5em{\scshape i\kern-0.25em b}\kern-0.8em\TeX}}}

\usepackage[colorinlistoftodos]{todonotes}
\usepackage{soul}

\newcommand{\REcomment}[1]{{\textcolor{blue}{\bf [Rudi: #1]}}}
\renewcommand{\REcomment}[1]{}

\newcommand{\PBcomment}[1]{{\textcolor{brown}{\bf [Parinaz: #1]}}}
\renewcommand{\PBcomment}[1]{}

\newcommand{\hide}[1]{}

\begin{document}

 \title{Exchanging Best Practices and Tools for Supporting Computational and Data-Intensive Research  \\ \LARGE The Xpert Network}

\author{Parinaz Barakhshan}
\affiliation{%
  \institution{University of Delaware}
  \streetaddress{Oklahoma State Drive}
  \city{Newark}
  \state{Delaware}
  \country{US}}
\email{parinazb@udel.edu}

\author{Rudolf Eigenmann}
\affiliation{%
  \institution{University of Delaware}
  \streetaddress{P.O. Box 1212}
  \city{Newark}
  \state{Delaware}
  \country{US}}
\email{eigenman@udel.edu}

\begin{abstract}

We present best practices and tools for professionals who support computational and data-intensive (CDI) research projects. The practices resulted from an initiative that brings together national projects and university teams that include individual or groups of such professionals. We focus particularly on practices that differ from those in a general software engineering context. The paper also describes the initiative -- the Xpert Network -- where participants exchange successes, challenges, and general information about their activities, leading to increased  productivity, efficiency, and coordination in the ever-growing community of scientists that use computational and data-intensive research methods.

\end{abstract}

\maketitle

\pagestyle{plain}

\section{Introduction}
\label{sec:Intro}

\subsection{Motivation and Key Contributions}
The motivation of this paper, and its underlying project, was that there appears to be little exchange of information between the increasing number of support groups for computational and data-intensive (CDI) research, which are of critical importance, as laid out below. In particular, knowledge of best practices is usually acquired by "training on the job" and not communicated to other projects. What's more, many domain science projects, by budgetary realities, include individual computational experts, who have little to no access to peers and need to self train on such practices. The fact that the practices exhibit differences from those in a general software engineering context makes their compilation and dissemination especially relevant. Filling this gap is the key opportunity and contribution of the present paper. We describe initial outcomes of an initiative that brings together CDI support professionals, which we refer to as CDI experts or, short, Xperts. There are many related terms, such as research software engineers (RSEs), research facilitators, or  research programmers. While one can argue about differences, in this paper we consider the terms essentially synonymous.


\subsection{Importance of Supporting Computational and Data-intensive (CDI) Research}
The importance of CDI research is well evidenced.  A relentlessly growing amount of computing power is consumed to conduct research using computational experimentation.  Nobel prizes have been awarded to CDI science breakthroughs~\cite{Day2012}. {\itshape Laszewski et al.} have shown that CDI work correlates with higher scientific impact~\cite{LWFH+15}. {\itshape Apon et al.} have found that Universities investing in computational infrastructure supporting CDI research tend to increase in ranking~\cite{AADG+10}.

Supporting such CDI research by Xperts is critical, allowing domain scientists to focus on their research tasks. Recognizing this need, several large projects funded by the National Science Foundation (NSF)~\cite{NSF}, such as XSEDE~\cite{xsede}, CyVerse~\cite{CyVerse}, and the NSF Software Institutes, support the community through groups of Xperts. There are also a growing number of university IT departments, or research computing support groups, that include Xperts for boosting the productivity of researchers on their campuses. Furthermore, many large domain-science projects include computational experts for supporting their own applications. The effectiveness of such computational support is remarkable.
A study by {\em XSEDE's ECSS group}~\cite{ecss} has revealed that time invested by computational experts can have a four-fold return in terms of time a domain scientist would spend on the same tasks.

\subsection{Structure and Activities of the Xpert Network}
\label{sec:Xpert}

The Xpert Network~\cite{xpert} is an NSF-supported, community-serving initiative that offers
\begin{itemize}
    \item Monthly online meetings (webinars and panels) to exchange best practices applied, tools used, and open problems faced in computational and data-intensive (CDI) research,
    \item Face-To-Face Meetings at major conferences, such as PEARC (Practice and Experience in Advanced Research Computing), ICS (International Conference on Supercomputing), and SC (Supercomputing) for in-depth discussions,
    \item A website (sites.udel.edu/xpert-cdi) with access to recordings and reports of past events, a calendar of upcoming events, and community announcements.
\end{itemize}

Under preparation are also 
\begin{itemize}
    \item A discussion platform for community communication around the clock, and
    \item An exchange program that supports participants visiting other participants.
\end{itemize}

The ultimate goal of the initiative is to advance science by increasing the productivity of researchers who use {\upshape CDI methods} for pushing research frontiers. To this end, several communities are invited to participate, including 

(i) researchers developing and using CDI applications, (ii) those who assist these researchers with expertise in CDI technology and methods (Xperts, Facilitators, RSEs), (iii) domain experts/scientists on university Campuses and other research organizations, and (iv) developers of tools that support the creation and use of CDI applications. These groups are invited to:

\begin{itemize}
    \item Share success stories about how Xperts have assisted domain researchers,
    \item Present open problems and challenges faced when supporting CDI research,
    \item Exchange best practices that are being applied in their organization or in their research,
    \item Present tools and reporting on the use of tools that have made a substantial difference in supporting the work of Xperts and domain researchers,
    \item Coordinate activities that are of mutual interest.
\end{itemize}

\subsection{Best Practices and Tools Guide}
\noindent Two specific outcomes of the  discussions are the creation of
\begin{itemize}
    \item A best-practices guide for computational and data-intensive research, and 
    \item A catalog of tools that support CDI research and engineering.
\end{itemize}
\noindent They will serve as training material for new Xperts joining CDI support teams, for instructional events, and for university curricula. The present paper can be seen as a starting point for such educational material. This material is related to the large volume of best practices for software engineering and scientific programming that exists, but differs in the following significant ways.

While training material for general software engineering and scientific programming cover important skills that {\em Xperts} must have, we are looking for best practices specific to professionals that {\em support domain science teams}. We focus on information that is not commonly found in software engineering and scientific programming literature, or on advice that is unexpected. We also discuss skills that our participants have called out as especially relevant and those that need to be applied differently in a {\em CDI context}. The sources of this information are the {\em Xpert Network webinars} that we have conducted thus far and three in-person events: the {\em ICS19 workshop}, and the {\em PEARC19 } and {\em SC19 Birds-of-a-feather sessions,} as well as our past work on several interdisciplinary projects with domain scientists.

Another important distinction of our approach is the inclusion of tools. Best practices need to be supported by tools that follow the underlying methodologies. Good tools for scientific software development can boost the productivity of domain science research as much as the best practices themselves.

\hide{
While some RSEs started as software developers and then learned about the research process, Other RSEs started as researchers who needed to write software to answer their research questions. The Xpert network aims to build common ground between these RSEs and improve their toolbox which will increase the productivity of CDI science and engineering. Learning best practices, and engaging with the broader software development community, RSEs help make research software more robust, manageable, and sustainable.  
}


The remainder of the paper is organized as follows. 
Section~\ref{sec:BestPractices} describes the best practices that have emerged in the Xpert Network activities so far. Section~\ref{sec:Tools}  focuses on tools that have proven their value in accelerating CDI research and the work of Xperts. Section~\ref{sec:Related}  describes related efforts and creating synergy among them, followed by conclusions in Section~\ref{sec:Conclusion}.

\mybox{Summary of Best Practices }{gray!40}{gray!10}{
\textbf{Best Practices for Collaborations: }
\justify
\begin{enumerate}

\item \textbf{Diversity of Xpert Backgrounds-} Be aware of different backgrounds Xperts may bring into a team; configure training so that less-familiar best practices can be acquired as needed.

\item \textbf{Understanding the Academic Environment-} Be aware of the academic reward system and activities to increase academic standing.

\item \textbf{Breadth of Xpert Skills Needed-} Prepare for skills needed beyond your current expertise, by networking with other Xperts.

\item \textbf{Collaborative Assistance-} Help propel new projects though short-term, close collaboration with domain scientists.

\item \textbf{Overcoming the Terminology Gap-} Carefully identify and resolve terminology gaps. Keep vocabulary to the essentials. Explain using many examples.
\end{enumerate}

\noindent\textbf{Best Practices for Software Development:}
\justify
\begin{enumerate}
\setcounter{enumi}{5}

\item \textbf{Developing a Project Plan-} Devote substantial time to understanding the domain problem, turning possibly vague ideas into a feasible solution approach, and developing a project plan.

\item \textbf{Prioritize Functional Requirements-} Carefully vet all requirements by the application’s end users and prioritize aggressively.

\item \textbf{Issue Tracking-} Track origin as well as implementation status of requirements and bug reports.

\item \textbf{Source Code Management-} Make use of source code management and version control systems to track you software’s evolution.

\item \textbf{Code Review-} Xpert and domain scientist should review each other’s software written. 

\item \textbf{Software Testing-} Define test cases that the application and its components must pass before you begin their implementation.

\item \textbf{Documentation-} Document your project to ensure long-term success, reproducibility, and obtain proper credit for your work.

\item \textbf{Continuous Integration-} Integrate new software updates frequently into the application version seen by the end users in their end environments.

\item \textbf{Reproducibility-} Enable reproducibility and transparency by capturing data and software underlying scientific processes, using available software platforms.

\item \textbf{Parallelization-} Write serial code first, then parallelize.
\end{enumerate}
}

\section{Best Practices}
\label{sec:Bestp}
\label{sec:BestPractices}

This section identifies best practices for CDI research involving applications that are {\em compute intensive} (spending most execution time on computational algorithms) and/or {\em data intensive} (manipulating large volumes of data). Two broad categories of best practices for supporting such research have emerged in the Xpert Network activities: (i) best practices related to project collaborations and (ii) best practices for software development. As mentioned in the introduction, the second category is covered well in the software engineering literature and in organizations focused on training, such as {\em Software Carpentry}~\cite{SBA+08}. 

This paper considers only those software development practices that have been described by participants as different or particularly important for CDI research. Of particular interest are practices that are {\em actionable}; we will highlight the recommended actions in each practice. In some cases, the action lies in {\em being aware} of something. The following subsection will begin with this category.

\subsection{Diversity of Xpert Backgrounds}
{\bf Be aware of different backgrounds Xperts may bring into a team; configure training so that less-familiar best practices can be acquired as needed.} Members of {\em Xpert groups} may come from domain sciences, from computer science backgrounds, or from environments where they learned to assist {\em CDI } science teams through training on the job.  For example, working with Unix commands, parallel programming models, and version control may be second nature to members with computer science background, whereas these very skills may be critical best practices to acquire by {\em Xperts} with domain-science background. Vice versa, {\em  Xperts} who have a domain degree are often well aware of the terminology gap discussed in Section~\ref{subsec:terminologygap}, which facilitates the communication with new science collaborators~\cite{ICS2019report}. Training for Xperts should accommodate these differences, allowing the trainees to focus on best practices they are not familiar with.

\subsection{Understanding the Academic Environment}
\label{subsec:academic}
{\bf Be aware of the academic reward system and activities to increase academic standing.}
An issue for {\em Xperts} that is essentially absent in a general software engineering context is the importance of understanding the academic environment. This issue is especially relevant for software engineers with background in industry, where hierarchical organization, and the overriding goal of creating a reliable product as rapidly as possible, are the norm.
Understanding the academic reward system and the many side activities that researchers may get engaged in to maintain the academic standing of the science team~\cite{virtualResidencyHenry,Virtualresidency} will influence project decisions. In fact, some of the best practices need to be understood from this view point, such as the issues of documentation and testing, mentioned in Sections ~\cref{subsec:documentation} ~\cref{subsec:testing}.

\subsection{Breadth of Xpert Skills Needed}
{\bf Prepare for skills needed beyond your current expertise, by networking with other Xperts.}
Modern CDI applications draw from a broad range of technologies, such as computing paradigms, programming languages,
architectures, and algorithms.
What's more, this range keeps evolving. For example, new application may demand expertise of machine learning techniques. Individual Xperts and those in  small teams, serving a large CDI research community, will unlikely cover the needed skills. Besides, they may need to perform tasks across the entire software life cycle and be involved in project management. 
It is important to be able to recognize such situations and seek external advice.  Maintaining contacts with other Xpert teams, who may be consulted when needed, is highly advisable. 



\subsection{Collaborative Assistance Between Xperts and Domain scientists}
\label{subsec:collaboration}
{\bf Help propel new projects though short-term, close collaboration with domain scientists.}
A recommended form of interaction between Xperts and CDI domain scientists is through  {\em collaborative assistance}. For a period of one to several months, Xperts work side-by-side (physically or virtually) with the domain scientists whose project they support. During the collaboration, the Xpert takes care of computer-engineering issues that arise, while the domain researcher resolves the problems requiring application science expertise. This form of joint work allows issues that fall into the competency of the collaborator to be addressed immediately, significantly reducing, or even avoiding, the need for {\em domain researchers and Xperts} to get trained on each others' knowledge and skills at project begin. Over the course of the joint work, the collaborators tend to pick up sufficient knowledge of each others' skills and terminology. In particular, the domain scientist becomes familiar with the software development processes and tools, allowing them to continue the work independently. Vice versa, the {\em Xpert } will have acquired sufficient knowledge of the domain problem, allowing them to effectively provide help remotely, even after moving on to other projects. This model has been pioneered by  {\em XSEDE's ECSS group} and was applied successfully by other Xpert teams~\cite{ICS2019report}.

\subsection{Overcoming the Terminology Gap Between Computer and Domain Sciences}
\label{subsec:terminologygap}
{\bf Carefully identify and resolve terminology gaps. Keep vocabulary to the essentials. Explain using many examples.}
The gap between computer-science and domain-science lingo is an often mentioned issue in Xpert Network discussions. The challenge can be big if the same term is used by both collaborators, but with different meanings. Participants reported significant confusions and even incorrect project executions due to this challenge. Awareness of the issue and patience in trying to understand the collaborators' viewpoints are critical. It was pointed out that the gap is wider than in a general software engineering context, when trying to understand a customer, as science terminology tends to use rich vocabularies. Keeping the vocabularies to the essentials and investing time to explain new terms is key to successful collaboration. Using many examples and frequent feedback from both sides will help bridge this gap. 

\subsection{Understanding the Domain Problem and Developing a Project Plan}
\label{subsec:understandingTheDomain}
{\bf Devote substantial time to understanding the domain problem, turning possibly vague ideas into a feasible solution approach, and developing a project plan.}
Many Xpert Network participants highlighted the importance of the first contact with the supported domain scientists and the approach taken to understand problems and develop solutions. In addition to awareness of the terminology gap, the relevance of investing time in understanding the goals and the functional as well as  non-functional requirements was emphasized. Showing patience in the process is critical. The domain researcher needs to be helped to transform an initially often vague idea into a concrete plan. Developing specific requirements for the computational application and the underlying system is essential. Devote sufficient time. One challenge for the Xpert is to introduce the researcher to the possibilities of the to-be-created or improved software implementation and the underlying system while introducing only a small number of new terms. Recommendations include the use of many concrete examples and an {\em inverted pyramid} approach. The latter describes a possible solution in initially very few, high-level terms, which are then successively refined in discussions.

The situation is again related to that of software engineers with their customers,  two important differences being: 
\begin{enumerate}
\item the potentially large terminology gap, mentioned above, and 
\item the collaborative assistance situation, which often  enables the inverted pyramid approach: As {\em Xpert} and domain researcher work side by side, refinements of high-level ideas can more easily happen over time, permitting the collaborators to invest their initial effort in defining the tip of the pyramid.
\end{enumerate}



\subsection{Prioritize Functional Requirements}
\label{subsec:prioritize}
{\bf Carefully vet all requirements by the application's end users and prioritize aggressively.} The dilemma of a large number of desirable features but only a short project duration can be  big in scientific software. Xperts need to learn to tell the difference between essential and desirable  requirements users may have. Essential features should be strictly prioritized. Scientific software is also special in that requirements can change rapidly, as new insights of a research project emerge. Frequent re-assessment of the requirements and priorities is needed ~\cite{SWLifeCycle}.

\subsection{Issue Tracking}
\label{subsec:tracking}
{\bf Track origin as well as implementation status of requirements and bug reports.} Issue tracking is important in most mid-size and large-size software development. Two points make it a particular concern in scientific applications. The first is the belief that research projects -- often of a three-year duration -- will remain small and easy to oversee, obviating the need for issue tracking. The second is that developers (typically graduate students and postdoctoral researchers) tend to change often, losing important project memory. The reality is that, even within three years, remembering what feature was requested by whom and with what rationale, can be difficult. After a "change of guard", the same will become close to impossible. What's more, successful science applications may turn out to have a long life. Hence, being able to trace a feature request, its implementation status, and the reasons for accepting/rejecting the request to its origin can be key. Furthermore, having a clear record of who made what request can be critical  for re-assessing and re-prioritizing functional requirements, mentioned previously.

Section~\cref{subsec:toolsIssueTracking} describes issue tracking tools, which maintain a list of tasks and sub-tasks to be performed, avoiding duplicated efforts and enabling collaborative work. The tools support requirement traceability, that is, the association of an application update with the requirement that motivated it~\cite{wilson2013software}.

\subsection{Source Code Management and Version Control}
\label{subsec:versionControl}
{\bf Make use of source code management and version control systems to track you software's evolution.}
An {\em Xpert} with software engineering background is well aware of the importance of software version control. At the same time, this importance needs to be advertised to domain experts, as pointed out strongly by the Xpert network participants.

Many successful science software applications had their origin in a {\em  "toy program"} written by a graduate student. It got gradually expanded by several authors, caught the attention of a wide audience, and ended up becoming an important research tool. Without version tracking, the history (origin, authors, relationship of features, and specific extensions) may no longer be known, making it difficult to extend further and obtain needed documentation. Learning version control methods and tools will not only help overcome these difficulties; it will also increase the productivity of the software developer as soon as the application exceeds the boundaries of a small program. What's more, source code management tools greatly facilitate collaborative software development~\cite{AMPgateway}, enable software roll-back to a previous, well-defined state, and help developers start a new branch of the software. The latter can be important if, say, two graduate students want to add their own, separate feature sets. Section~\cref{subsec:toolsSourceCodeManagement} will mention supporting tools.

\subsection{Code Review}
\label{subsec:codeReview}
{\bf Xpert and domain scientist should review each other's software written.}
Code review -- a second person reading newly written software -- is one of the best ways to catch errors and also to improve the code. Despite its effectiveness, it is often ignored, unless strictly mandated, and hence warrants mentioning here. A number of additional reasons make this practice particularly useful for an Xpert-domain scientist relationship: 
\begin{itemize}
\item  The two collaborators have different backgrounds, increasing the chance of detecting an issue that eluded the other.
\item  The review is effective in verifying that requirement and implementation match.    
\item  It is a good way of transferring the knowledge of what the Xpert did to the domain scientist, who will continue the work after the collaboration completes.
\item The discussions happening during code review provide excellent material for documenting the code and project.
\end{itemize}

\noindent Code review helps improve the science as well as the code itself. Research software can benefit from code reviews, as much as industrial software~\cite{codereview}.

\subsection{Software Testing}
\label{subsec:testing}
{\bf Define test cases that the application and its components must pass before you begin their implementation.} Testing is another issue that is covered well in the software engineering literature. The topic has been mentioned often in our Xpert discussions, deserving a place in this list of best practices. Next to raising awareness of the need and the benefit of testing, a key point is that  testing must not be an afterthought;  one must avoid thinking of test cases only when the software is close to completion. Instead, testing should be considered in the design phase. In the initial communication with the domain researcher, the Xpert needs to ask the question "what test cases should be passed by the tool or application, once it is completed?" Coming up with a thorough set of such cases will not only facilitate the later testing phase; it will also help clarify the specific capabilities that need to be implemented. The implementation of features that do not satisfy any test case, can be postponed or even avoided altogether.
Designing test cases can go hand in hand with functional prioritization (\cref{subsec:prioritize}).

\subsection{Documentation}
\label{subsec:documentation}
{\bf Document your project to ensure long-term success, reproducibility, and obtain proper credit for your work.}
Software engineering teachings cover extensively the fact that well documented programs are essential for software maintenance and extensibility. Yet, academic software is notorious for the lack of documentation, as pointed out forcefully by {\em Xpert Network } participants. While raising awareness of the benefits is important, one also has to understand that academic processes and reward systems often do not support spending time on documentation (\cref{subsec:academic}). Most academic research funding pays for showing principles and developing prototypes, not production-ready software. What's more, students are under pressure to graduate, whereby the  functionality of the created software is more important than its documentation. If the software becomes successful, the original author often is no longer involved and thus has little incentive to retrofit a proper description. It is often the difficult task of the next generation of students in a team to accomplish these tasks. It helps, however, to understand these relationships. If one can identify the original creator, they may be willing to help, especially when offering them proper credit or co-authorship on a forthcoming publication. Vice versa, properly documenting the original authorship will insure that such credit can be given.

Reproducibility (\cref{subsec:reproducibility}) of scientific research is a concern that is currently gaining attention. Proper software documentation can be key to reproducibility~\cite{AMPgateway}. 

\subsection{Continuous Integration}
\label{subsec:CI}
{\bf Integrate new software updates frequently into the application version seen by the end users in their end environments.} Continuous Integration (CI) is generally good software engineering advice. It catches miscommunicated requirements early. CI also allows users to experience new features and provide feedback to the developer early and often. The process may be combined with automated builds and testing, ensuring that new functionality satisfies and continues to satisfy defined test cases. 

CI was mentioned as particularly relevant for the work of Xperts, as scientific software tends to be a moving target with frequent changes of requirements. Ensuring that new features are what the end user had in mind, conform to defined test cases, and do not break previous requirements is critical. Many teams find that this approach leads to significantly reduced integration problems and enables the development of cohesive software more rapidly.

\subsection{Reproducibility}
\label{subsec:reproducibility}
{\bf Enable reproducibility and transparency by capturing data and software underlying scientific processes, using available software platforms.}
While the reproducibility of research results is a key requirement in any domain of science, there is recent, increased focus on this issue in computation-based research. Reproducibility was highlighted as a significant concern in our Xpert Network discussions as well, pointing out that adding  supporting software and data to a publication can increase the value significantly. This is of particular importance in large computational studies, where data analysis may play a central role in reaching the conclusions. Disclosing the data and software underlying the research methods will add transparency.

Making use of software platforms, such as GitHub~\cite{GitHub}, GitLab~\cite{GitLab}, and container environments, such as Docker~\cite{docker}, can dramatically reduce the cost of capturing and describing the computing environment used to produce the scientific results.

\subsection{Parallelization}
\label{subsec:parallelization}
{\bf Write serial code first, then parallelize.}
The question if parallel code should be written directly, versus creating serial code first, is an open one~\cite{ParallelProg}.  The Xpert Network participants were clear, however, in recommending the latter for creating
Computational and Data-Intensive (CDI) applications. Among the arguments were that the benefits of lesser complexity of getting the serial code correct first, combined with better tool support for serial code, outweigh the negatives. The primary negative is that certain serial algorithms are intrinsically hard to parallelize. When keeping this issue in mind and selecting algorithms that are known to parallelize and scale well, this negative can be overcome, however.

\mybox{Summary of Tools }{gray!40}{gray!10}{
\textbf{Tools for CDI Application Development: }
\justify
\begin{enumerate}

\item \textbf{Project Management --} Jira~\cite{Jira}, Kanban boards~\cite{kanbanboards}

\item \textbf{Documentation --} Doxygen~\cite{Doxygen} (for C, C++, C\#, D, Fortran, Java, Perl, PHP, Python), GhostDoc~\cite{GhostDoc} (for C\#, Visual Basic, JavaScript), Javadoc~\cite{Javadoc} (for Java).

\item \textbf{Source Code Management --} Git~\cite{Git}, GitHub~\cite{GitHub}, GitLab~\cite{GitLab}, Bitbucket~\cite{Bitbucket}, Mercurial~\cite{Mercurial}.

\item \textbf{Issue Tracking --} 
Jira~\cite{Jira}, Trello~\cite{trello}, Github Boards~\cite{GitHubBoards}, Asana~\cite{asana}

\item \textbf{System Build--} CMake ~\cite{CMake}, GNU Make~\cite{GNUMake}

\item \textbf{Compiler Reports and Diagnostics --} Intel~\cite{IntelCompilers}, Gnu~\cite{GNUCompilers}, PGI~\cite{PGICompilers}.
Research Compilers: Cetus~\cite{Cetus}, Rose ~\cite{Rose} 

\item \textbf{Debuggers --} GDB (GNU Project debugger)~\cite{GDB}, Arm DDT ~\cite{ArmDDT}

\item \textbf{Memory Debuggers --} Valgrind~\cite{Valgrind}, AddressSanitizer (ASan)~\cite{AddressSanitizer}

\item \textbf{Performance Analysis --} Intel Advisor~\cite{IntelAdvisor}, ARM Map~\cite{ArmMAP}, Tau~\cite{TAUperformance}, HPCtoolkit~\cite{HPCToolkit}, mpiP~\cite{mpiP}

\item \textbf{Test Frameworks --} Reframe test framework~\cite{ReFrame}  

\item \textbf{Containers --} Docker~\cite{docker}, Singularity~\cite{Singularity} 

\item \textbf{Cloud-based Development Environments --}Eclipse Che~\cite{EclipseChe}, Amazon Cloud9~\cite{AmazonCloud9}, Gitpod~\cite{Gitpod}, Codespaces~\cite{codespaces} 
\item \textbf{Continuous Integration --} Travis CI~\cite{travisCI}, GitLab~\cite{GitLab}, Jenkins~\cite{jenkins} 

\item \textbf{Profiling/Tracing --} GNU Project Profiler (GPROF)~\cite{gprof}, TAU~\cite{TAU}

\item \textbf{User interfaces to HPC resources --} 
\begin{itemize}
\item {Science Gateways~\cite{scienceg}}
\item{Open OnDemand ~\cite{openondemand} }
\item{Rich desktop clients, such as the Eclipse Parallel Tools Platform (PTP) ~\cite{EclipsePTP}}
\item{Interactive applications, such as Jupyter~\cite{jupyter}, RStudio~\cite{rstudio}, JupyterHub~\cite{jhub}, and JupyterLab~\cite{jupyter}}
\end{itemize}  
\end{enumerate}
}

\section{Tools for Accelerating CDI research}
\label{sec:Tools}
This section summarizes discussions of CDI application development tools in the Xpert Network workshops, webinars, and position papers.
The importance of such tools for increasing the productivity of both CDI domain scientists and those who assist them was pointed out repeatedly. Many of the best practices can be enhanced by supporting tools, for example, by identifying opportunities for optimization and parallelization, automating performance analysis and testing, or by supporting the workflow, such as issue tracking and version control. Listed below are some of these tools that have had an impact in developing CDI applications of Xpert Network participants~\cite{ICS2019report}. 



\subsection{Project Management}
\label{subsec:toolsProjectManagement}
Many groups report good experiences using tools for managing project tasks. Use of such tools help with  visualizing the work, limits work-in-progress, helps teams establish order in their daily work, and maximizes efficiency. These tools also help facilitate communication between groups of collaborators, such as between domain researchers and computational experts.

\noindent\textbf{Tools:} Jira~\cite{Jira}, Kanban boards~\cite{kanbanboards}

\noindent\textbf{Best practices supported by these tools:} Collaborative Assistance (\cref{subsec:collaboration}), Developing a Project Plan (\cref{subsec:understandingTheDomain}), Prioritize Functional Requirements (\cref{subsec:prioritize}), Issue Tracking (\cref{subsec:tracking})

\subsection{Documentation}
\label{subsec:toolsDocumentation}
The challenge of scientific software documentation was mentioned in Section~\cref{subsec:documentation}. Researchers tend to write comprehensive documentation only when absolutely demanded by a collaborator or external user of the code.  Tools can help overcome this issue.

\noindent\textbf{Tools:} Tools that help in creating and automating software documentation are Doxygen~\cite{Doxygen} ( for C, C++, CSharp, D, Fortran, Java, Perl, PHP, Python ), GhostDoc~\cite{GhostDoc} ( for CSharp, Visual Basic, JavaScript ), and Javadoc~\cite{Javadoc} (Java). 
Among the tools that can be used for the sole purpose of publishing documentation, GitHub and GitHub Pages were mentioned.

\noindent\textbf{Best practices supported by these tools:} Documentation(\cref{subsec:documentation})

\PBcomment{Added more Description}
\subsection{Source Code Management}
\label{subsec:toolsSourceCodeManagement}
Tools for source code management and version control were described as fundamental to the development of CDI applications. At the same time, participants reported that some domain teams used ad-hoc methods, instead.
The following tools are especially important where multiple developers are continuously involved in changing the source code.

\noindent\textbf{Tools:} Git~\cite{Git}, GitHub~\cite{GitHub}, GitLab~\cite{GitLab}, and Bitbucket~\cite{Bitbucket}, Mercurial~\cite{Mercurial}

\noindent\textbf{Best practices supported by these tools:} Source Code Management and Version Control (\cref{subsec:versionControl})

\subsection{Issue Tracking}
\label{subsec:toolsIssueTracking}
Issue tracking systems manage and maintain lists of issues, recording implementation status and traceability to user requests. The tools are generally used in collaborative settings but can also be employed by individuals.

\noindent\textbf{Tools:} Jira~\cite{Jira}, Trello~\cite{trello}, Github Boards~\cite{GitHubBoards}, and Asana~\cite{asana}

\noindent\textbf{Best practices supported by these tools:} Issue Tracking (\cref{subsec:tracking})

\subsection{System Build }
\label{subsec:buildsystem}
System build tools support the generation of executables and other translated files from the program’s source code, and software packaging.~\cite{BuildSystems}

\noindent\textbf{Tools:} CMake ~\cite{CMake}, GNU Make~\cite{GNUMake}

\noindent\textbf{Best practices supported by these tools:} Continuous Integration (\cref{subsec:CI}),
Parallelization (\cref{subsec:parallelization})

\subsection{Compiler Reports and Diagnostics}
\label{subsec:CompilerReports}
Besides their code-generation functionality, compilation tools can be crucial in providing reports about analyzed program characteristics and applied optimizations.

\noindent\textbf{Tools:} Intel~\cite{IntelCompilers}, GCC ( GNU Compiler Collection )~\cite{GNUCompilers}, PGI~\cite{PGICompilers}, LLVM~\cite{llvm}.
Research Compilers: Cetus~\cite{Cetus}, Rose ~\cite{Rose}, 

\noindent\textbf{Best practices supported by these tools:} Parallelization (\cref{subsec:parallelization})

\subsection{Debuggers}
\label{subsec:debuggers}
The following are among the debugging tools mentioned. Attendees reported on the difficulty of finding convenient  tools that assist in debugging parallel programs.

\noindent\textbf{Tools:} GDB ( GNU Project debugger )~\cite{GDB}, Arm DDT ~\cite{ArmDDT}

\noindent\textbf{Best practices supported by these tools:} Parallelization (\cref{subsec:parallelization})

\subsection{Memory Debuggers}
\label{subsec:MemDebuggers}
Memory debuggers have been useful for tasks such as identifying  memory leaks and buffer overflows, often related to  allocation and deallocation of dynamic memory. 

\noindent\textbf{Tools:} Valgrind~\cite{Valgrind}, AddressSanitizer (ASan)~\cite{AddressSanitizer}

\noindent\textbf{Best practices supported by these tools:} Parallelization (\cref{subsec:parallelization})

\subsection{Performance Analysis}
\label{subsec:PerfAnalysis}
Performance analysis tools have been essential for understanding, diagnosing, visualizing and resolving issues related to program performance and scalability. 

\noindent\textbf{Tools:} Intel Advisor~\cite{IntelAdvisor}, Arm Map~\cite{ArmMAP}, TAU~\cite{TAUperformance}, HPCtoolkit~\cite{HPCToolkit}, mpiP~\cite{mpiP}

\noindent\textbf{Best practices supported by these tools:} Parallelization (\cref{subsec:parallelization})

\subsection{Test Frameworks}
\label{subsec:TestFrameworks}
Testing frameworks provide guidelines and rules for creating and designing test cases. They can provide essential support in increasing the efficiency of software testing. 

\noindent\textbf{Tools:} Reframe test framework~\cite{ReFrame}

\noindent\textbf{Best practices supported by these tools:} Software Testing (\cref{subsec:testing})

\subsection{Containers}
\label{subsec:toolsContainers}
 Containerization helps preserve software environments and addresses reproducibility. Containers can capture the full software environment of a computational method and enable portability to different platforms. It was mentioned that, while containers do not solve the issue of long-term preservation of software -- due to changing container versions, operating systems, and research infrastructures -- they may deliver intermediate solutions~\cite{ScienceGateways}.

\noindent\textbf{Tools:} Docker~\cite{docker} and Singularity~\cite{Singularity}

\noindent\textbf{Best practices supported by these tools:} Reproducibility (\cref{subsec:reproducibility})

\subsection{Cloud-based Development Environments}
\label{subsec:toolsCloudEnv}
These tools  help with collaborative software development, documentation, debugging and testing. They provide convenience by providing access via a browser, without the need for download and installation. Many open challenges were mentioned, however. They relate to the support for real time co-development, co-debugging problems in the same session, and reproducibility. The cost of running cloud-based services, security, privacy, and the complexity of the application building and execution process were also mentioned as concerns~\cite{cloudIDE}.

\noindent\textbf{Tools:} Eclipse Che~\cite{EclipseChe}, Amazon Cloud9~\cite{AmazonCloud9}, Gitpod~\cite{Gitpod}, and Codespaces~\cite{codespaces}

\noindent\textbf{Best practices supported by these tools:} Collaborative Assistance (\cref{subsec:collaboration}), Documentation (\cref{subsec:documentation}), Software Testing (\cref{subsec:testing})

\subsection{Continuous Integration}
\label{subsec:toolsCI}
 Cloud-based access to tools for continuous integration, can significantly improve development productivity and reduce the amount of human involvement for routine setup and maintenance tasks. In this way, testing is triggered on cloud servers when commits are pushed to the source code repository~\cite{cloudIDE}.

\noindent\textbf{Tools:} Travis CI~\cite{travisCI}, GitLab~\cite{GitLab},  and Jenkins~\cite{jenkins}

\noindent\textbf{Best practices supported by these tools:} Continuous Integration (\cref{subsec:CI})

\subsection{Profiling/Tracing}
\label{subsec:toolsProfiling}
 One of our surprise findings was that, even though profiling is a technology well known to software engineers, many domain scientists were insufficiently aware of it. After introducing them to the basic concepts and features, they reported making significant progress in analyzing and improving their programs.

\noindent\textbf{Tools:} Beyond profiling functionality of basic tools, such as GPROF ( GNU Project Profiler ) ~\cite{gprof}, the TAU~\cite{TAU} environment offers a rich feature set for both profiling and program tracing to identify performance bottlenecks. TAU's automatic instrumentation capabilities support programs written in many languages and parallel programming models.

\noindent\textbf{Best practices supported by these tools:} Parallelization (\cref{subsec:parallelization})

\subsection{User interfaces to HPC resources}
\label{subsec:toolsUserInterfaces}
Accessing HPC machines through means other than command line interfaces are key for broadening participation~\cite{UIforHPC}. 

\noindent\textbf{Tools:} Among the approaches are:
\begin{itemize}
\item {Science Gateways~\cite{scienceg} are domain-specific web portals, often with access to high-performance computing resources.}
\item{Open OnDemand ~\cite{openondemand} provides comprehensive web interfaces with rich functionality for HPC and flexible customization for resource managers.}
\item{Rich desktop clients, such as the Eclipse Parallel Tools Platform (PTP) ~\cite{EclipsePTP}, provide  multi-language development environments. They include full support for multiple parallel paradigms including MPI, OpenMP, and OpenACC, as well as synchronized projects for effective utilization of remote HPC resources. PTP also provides rich customization for resource managers.}
\item{Interactive applications, such as Jupyter~\cite{jupyter} and RStudio~\cite{rstudio}, and their web interfaces JupyterHub~\cite{jhub} and JupyterLab~\cite{jupyter}, provide remote access to HPC resources.}
\end{itemize}  
\textbf{Best practices supported by these tools:} Parallelization (\cref{subsec:parallelization})

A general concern expressed with many of these tools was that improvements are needed to achieve both ease of use for end users and advanced functionality for power users~\cite{Pearc19Bof}. 


\section{Related Efforts and Synergy}
\label{sec:Related}
\subsection{Related Work on Best Practices}
A number of papers have recommended practices for scientific programming. 

{\em Wilson et al.}~\cite{wilson2014best, wilson2017good} have provided a set of "best practices" and also "good enough practices" for scientific computing community from the experiences of the thousands of people who have taken part in Software Carpentry~\cite{softwarecarpentry} and Data Carpentry~\cite{datacarpentry} workshops, and from a variety of other guides.

{\em Dubois} ~\cite{dubois1999ten} has described some of his experiences to help writing scientific programs. 

{\em Heroux et al.}~\cite{heroux2009barely} discuss practices used in the Trilinos ~\cite{trilinos} open-source software library project, some of which are close to practices advocated by the Agile software development community~\cite{Agile}.

{\em Naguib et al.}~\cite{naguib2010position} present an overview of two projects and describes the software engineering methods they applied on these computational science and engineering applications.

{\em Wilson}~\cite{softwarecarpentry} talks about the “common core” of modern software development that is taught through software carpentry courses to help computational scientists meet the standards that experimental scientists have taken for granted~\cite{wilson2006software}.

{\em Riley}~\cite{anlgov} advocates bringing knowledge of useful software engineering practices to HPC scientific code developers. While not prescribing specific practices, she emphasizes adopting practices that {\em help productivity}.

{\em The Better Scientific Software (BSSw)}~\cite{BSSW}  community is focused on software development, and sustainability that leads to improved software for computational science and engineering (CSE) and related technical computing areas.

{\em The Working Towards Sustainable Software Science: Practice and Experiences (WSSSPE)} ~\cite{WSSSPE} community meet frequently to discuss the challenges of software for science.

{\em The Software Engineering for Computational Science (SE4Science) } ~\cite{SE4Science} community is working to understand the differences, necessities, impacts, and barriers of applying general software engineering practices to research software So that scientific communities can adopt good software engineering practices.

{\em The UK Software Sustainability Institute}~\cite{SoftwareSustainabilityInstitute} has worked on facilitating the advancement of software in research by cultivating more sustainable research software.

{\em IDEAS-ECP }~\cite{IDEASECP} community addresses issues of software productivity and sustainability in the  Exascale Computing Project (ECP)~\cite{ECPProject}. 
 
{\em The UCAR ~\cite{UCAR} Software Engineering Assembly (SEA)~\cite{SEA}} is a community for software engineering professionals within UCAR, that addresses the issue of effective software engineering throughout UCAR.
 

\REcomment{The following is a key sentence. Is needs improvement. One can read this as if our practices just repeated already existing papers. This would not be acceptable for publication. We need to be more precise about HOW we differ.}
While these contributions provide important, general software engineering advice for science and engineering applications, the focus in this paper was on practices  that are specific to computational Xperts (a.k.a. Research Facilitators or Research Software Engineers) that support domain scientists. We reported best practices discussed by participants of the Xpert Network activities, which represent both large projects and individuals involved in such support roles.
\PBcomment{newly added}
We focused on the best practices related to project collaborations between  Xperts and  domain scientists, which is the essence of such projects, and the best practices for software development that are particularly important or should be applied differently in CDI research. 

It is worth mentioning that even for those practices that are in common between CDI reserach and general software engineering, the priority and importance of the practices differ considering different scopes defined for the projects based on different environments (Academic Environment, or Industry) these projects are designed for. While in academic environment a working prototype would be enough to prove a point, in industry we are trying to make a reliable production quality application.

\subsection{Related Work on Tools}
Many CDI-related tools are introduced and taught through platforms, such as the Carpentries~\cite{carpentry} and the HPC University~\cite{hpcuniversity}.

The Carpentries include Software Carpentry~\cite{softwarecarpentry},  Data Carpentry~\cite{datacarpentry}, and HPC Carpentry~\cite{hpccarpentry}. While Data Carpentry aims to teach fundamental concepts, skills and tools for working more effectively with data, Software Carpentry has been teaching researchers general computing skills and related tools. HPC Carpentry's search engine on the other side searches upon the resources of {\em HPC University (HPCU)}~\cite{hpcuniversity}, and {\em Computational Science Education Reference Desk (CSERD)}~\cite{CSERD} to help researchers with a broad range of topic in HPC and computational science,including tools.
\PBcomment{added new efforts. Can you think of any other effort that should be added here? Rose Compiler?}

Some of the tools that were brought up in the Xpert Network Webinars and panels are mentioned below:

The {\em Extreme-scale Scientific Software Stack (E4S)}~\cite{E4S} provides open source software packages for developing, deploying and running scientific applications on high-performance computing (HPC) platforms. E4S provides from-source builds and containers of a broad collection of HPC software packages.

{\em TAU Performance System®}~\cite{TAUperformance} is a portable profiling and tracing toolkit for performance analysis of parallel programs written in Fortran, C, C++, UPC, Java, Python.

{\em XDMoD Tool}~\cite{XDMoDTool}, which stands for XD Metrics on Demand, provides a wide range of metrics pertaining to resource utilization and performance of high-performance computing (HPC) resources. 

{\em HPCToolkit}~\cite{HPCToolkit} introduces an integrated suite of tools for measurement and analysis of program performance on computers ranging from multicore desktop systems to the nation's largest supercomputers. HPCToolkit supports measurement and analysis of serial codes, threaded codes (e.g. pthreads, OpenMP), MPI, and hybrid (MPI+threads) parallel codes. 


While there is overlap with the tools presented in the above projects, this paper reported on tools that were discussed by participants of the Xpert Network activities as relevant to their work in supporting CDI domain science groups.
\subsection{Related Efforts in Networking Xpert Groups}

The Xpert Network is related to and collaborates with a number of activities that involve or facilitate idea exchanges among professionals supporting computational and data-intensive (CDI) research.

The {\em Research Software Engineer Association (US-RSE)}~\cite{US-RSE} is an
initiative to create synergy among research software engineer individuals and groups at US universities and other institutions. This is a recent initiative with aims that are similar to those of the Xpert Network, and the two efforts seek to combine forces.

The {\em CaRCC (Campus Research Computing Consortium) effort}~\cite{CaRCC} also aims to engage a similar community as  US-RSE, with the professionalization (professional development and advocacy) of those involved in campus research computing being a particular focus.  CaRCC uses the term  Research Facilitator, which is closely related to the terms Xpert  and RSE.

The {\em Virtual Residency}~\cite{Virtualresidency} program focuses on training Xperts/ RSEs/ Research Facilitators.

The {\em XSEDE (The Extreme Science and Engineering
Discovery Environment)}~\cite{xsede} project includes two thrusts that support CDI application researchers. The  {\em ECSS group (Extended Collaborative Support Service)} ~\cite{ecss} has a large number of computational and data experts who work collaboratively with domain scientists in improving their applications. The {\em Campus Champions}~\cite{CampusChampions} program engages participants at many university campuses to raise awareness of  XSEDE services and help local users accelerate CDI research.

 The {\em CyVerse project}~\cite{CyVerse}, supporting data-driven discovery, includes community support services similar to that of XSEDE's ECSS group.

 Related support groups are also part of the NSF Software Institutes, including the {\em Science Gateway Community Institute}~\cite{SGCI} and {\em the Molecular Sciences Software Institute}~\cite{MolSSI}.

The {\em Xpert Network} exchanges ideas with all the above efforts. Best practices and toos identified in this paper may provide training material for the mentioned projects. In addition to presenting best practices, the agenda of monthly Xpert Network webinars include discussions of coordination among the many involved activities, with the overall goal of increasing efficiency in the science community.

\section{Conclusion}
\label{sec:Conclusion}
This paper presented best practices and tools used by those who support domain researchers in creating, improving, and running computational and data-intensive (CDI) applications. We refer to these professionals as computational experts, research software engineers, research facilitators or, short, Xperts. The information in this paper emerged from discussions and reports of the Xpert Network project, also presented in the paper. While both the best practices and the tools have overlap with those presented in general software engineering courses and literature, we have emphasized aspects in which the work of Xpert professionals differs. Furthermore, the paper discussed related projects that support CDI research and how to create synergy within this community.

\begin{acks}

This work was supported in part by the National Science Foundation under Award No. OAC-1833846.

\end{acks}

\balance
 
\bibliographystyle{ACM-Reference-Format}

\bibliography{acmpaper,my}

\appendix

\end{document}